\documentclass[11pt]{article}

\usepackage[english]{babel}
\usepackage[utf8]{inputenc}

\usepackage[utf8]{inputenc}
\usepackage[T1]{fontenc}
\usepackage{moreverb,url}

\usepackage[colorlinks,bookmarksopen,bookmarksnumbered,citecolor=red,urlcolor=red]{hyperref}

\usepackage{algorithm} 
\usepackage{algorithmic}

\usepackage{multirow}
\usepackage{colortbl}
\usepackage{array,booktabs}
\usepackage{xcolor}
\usepackage{makecell}
\usepackage{pifont}

\makeatletter
\def\BState{\State\hskip-\ALG@thistlm}
\makeatother

\usepackage{amssymb, authblk}
\usepackage{amsmath, bbm}
\usepackage{fullpage}

\usepackage{graphicx}
\usepackage{verbatim}

\usepackage{threeparttable}
\usepackage{algorithm}

\usepackage[usestackEOL]{stackengine}

\DeclareMathAlphabet{\mathpzc}{OT1}{pzc}{m}{it}
\usepackage{cleveref}
\usepackage{hyperref}[]
\hypersetup{
	colorlinks=true,
	linkcolor=blue,
	filecolor=magenta,      
	urlcolor=cyan,
	citecolor=blue,
}

\usepackage{xspace}

\usepackage{amsmath}
\usepackage{amsfonts}
\usepackage{amssymb}
\usepackage{graphicx}
\usepackage{hyperref}
\usepackage{moreverb,url,multirow, tabularx}
\usepackage[usestackEOL]{stackengine}

\allowdisplaybreaks
\newcommand{\norm}[1]{\left\lVert#1\right\rVert}

\usepackage{pgfplotstable}
\usepackage{booktabs}
\usepackage{xspace}

\title{Distributional data analysis of accelerometer data from the NHANES database using nonparametric survey regression models}

\author{Marcos Matabuena$^{1*}$, Alexander Petersen$^{2}$\\
 $^{1}$CiTIUS (Centro Singular de Investigaci\'{o}n en Tecnolox\'{i}as Intelixentes), Universidade de Santiago de Compostela, Spain \\  $^{2}$Department of Statistics, Brigham Young University \\
	$^{*}$\url{marcos.matabuena@usc.es}
	\\}

\date{\today}

\begin{document}
\maketitle

\begin{abstract}

Accelerometers enable an objective measurement of physical activity levels among groups of individuals in free-living environments, providing high-resolution detail about physical activity changes at different time scales. Current approaches used in the literature for analyzing such data typically employ summary measures such as total inactivity time or compositional metrics. However, at the conceptual level, these methods have the potential disadvantage of discarding important information from recorded data when calculating these summaries and metrics since these typically depend on cut-offs related to exercise intensity zones chosen subjectively or even arbitrarily. Furthermore, much of the data collected in these studies follow complex survey designs. Then, using specific estimation strategies adapted to a particular sampling mechanism is mandatory. The aim of this paper is two-fold. First, a new functional representation of a distributional nature accelerometer data is introduced to build a complete individualized profile of each subject's physical activity levels. Second, we extend two nonparametric functional regression models, kernel smoothing and kernel ridge regression, to handle survey data and obtain reliable conclusions about the influence of physical activity in the different analyses performed in the complex sampling design NHANES cohort and so, show representation advantages.

\end{abstract}

\section{Introduction}

A patient’s physical activity levels are an influential causal factor associated with the development of chronic diseases, mortality, life-span, and increased medical costs \cite{pedersen2006evidence,almeida2014150,bolin2018physical}. At the same time, regular physical exercise is one of the most effective interventions to control glucose values in diabetic patients \cite{mendes2016exercise}, reduce weight \cite{franks2017causal}, minimize the effects of aging \cite{sattler2020current}, and improve health in general \cite{friedenreich2021physical, burtscher2020run}, often without introducing pharmacological treatment. Most medical guidelines recommend $150$ minutes per week of aerobic exercise for the general population \cite{mendes2016exercise}. However, to ensure the intervention’s success, a personalized training prescription and evaluation are required \cite{matabuena2019prediction,buford2013toward}.   

Traditionally, in epidemiological studies, physical activity levels in the general population have been measured using methods that introduce subjectivity, such as surveys, sleep-logs, and daily diaries \cite{sirard2001physical}. Similarly, in professional sports, subjective assessment metrics such as the rate of perceived exertion (RPE) \cite{eston2012use} have been widely used.  With the boom of digital medicine \cite{kvedar2016digital} and the possibilities of monitoring patients in real-time through biosensors, the objective measurement of physical activity levels with these technologies is becoming increasingly common \cite{matabuena2019improved}. The estimation of energy expenditure using accelerometers is probably the most general and reliable procedure for this purpose at the moment.

Accelerometer data provides a vast source of information that quantifies the intensity, volume, and direction of physical activity in real-time in the period in which the device is worn. For the last $15$ years, multiple epidemiological studies have used these devices to infer physical activity patterns in various cohorts. For example,  in \cite{troiano2008physical}, the authors describe the physical activity patterns in the American population using simple summary measures by age-groups; in \cite{goldsmith2016new}, the authors analyze how physical activity patterns vary minute-to-minute through functional data  analysis techniques with children from  New York. Other studies use accelerometers to resolve complex questions such as the relationships between physical activity levels and short-term mortality or life-span \cite{lynch2010objectively,ekelund2019dose, tarp2020accelerometer}. Precisely in this domain, a remarkable recent study \cite{Wsmirnova2019predictive} showed that physical activity patterns may predict  mortality more accurately than well-established epidemiological variables such as age, smoking, and the presence of cancer. Answering these questions with precision is essential to guide public health policies and design physical activity routines that optimize the population’s health \cite{raichlen2020sitting, ding2020towards}. The National Health and Nutrition Examination Survey (NHANES) is a public database containing information on the American population’s physical activity levels during the period 2003-2006, and is the best-known database of accelerometers. Other cohorts with available accelerometer data include the Baltimore longitudinal study \cite{nastasi2018objectively}, and more recent studies with the UK Biobank \cite{strain2020wearable}  or the International Children’s accelerometer Database \cite{sherar2011international}, and have provided new clinical knowledge with different study populations and other or similar sampling designs.

In the current era of precision medicine \cite{kosorok2019precision}, these devices are also beneficial for individualized prescription of physical exercise, given that the data obtained is vital for the control and measurement of exercise performed in general and sports populations. For these reasons, accelerometer technology has also been gradually used to evaluate interventions and more beneficial physical activity therapies in clinical trials \cite{napolitano2010accelerometer}.  

From a statistical point of view, the analysis of this data is usually complicated, and summary measurements must be used to compress the information recorded by the curves obtained with these devices. One of the main methodological obstacles that must be overcome is that the curves can have different lengths, and the subjects are not in standardized conditions, so a direct time series or functional data analysis is not usually workable. Given the inherent difficulty of direct examination of this data, practitioners often define several target zones and quantify the proportion of time (or total time) that the individual spends in each target zone when the device is worn. In many domains, such as diabetes, these metrics are commonly referred to as time in range metrics \cite{beck2019validation, biagi2019individual, dumuid2018compositional} . When the characteristic vector obtained is a ratio-vector, several papers have recently been suggested to use specific compositional data analysis techniques \cite{dumuid2018compositional, dumuid2019compositional, doi:10.1177/0962280218808819}. Naturally, time-in-range metrics suffer from a loss of information, as the information is discrete in intervals. In addition, the cut-off points chosen may be arbitrary and dependent on the characteristics of the population under study.

In a recent work \cite{matabuena2020glucodensities}, the limitation of time in range metrics in diabetes with continuous glucose monitoring data (CGM) was overcome by a more comprehensive distributional representation of the data. Specifically, a generalization of the time in range metric approach was introduced and consisted of constructing a functional profile called a glucodensity that can be used to extract the proportion of time spent in any interval of glucose concentration levels. In addition, a framework of nonparametric statistical techniques was proposed based on the Wasserstein distance between glucodensities. The results from this application to diabetes showed  better clinical sensitivity compared to traditional approaches that include time in range metrics with cut-offs recommended by the American Diabetes Association \cite{battelino2019clinical}.

In this paper, a similar strategy to that of glucodensities is applied, exploiting the connection between the Wasserstein geometry and quantile functions, with the dynamic accelerometer data being represented as probability distributions. Here, the induced probability distribution differs from the glucodensity extracted from CGM data in that it is a mixed distribution containing an atom at zero representing the proportion of inactive time.  With the intended purpose of drawing more representative conclusions of physical activity levels at the population level than can typically be achieved with observational studies, many of the main cohorts' physical activity studies are designed with a complex survey structure including demographical characteristics in sample selection, as is notably the case in the NHANES database. In order to handle these sampling characteristics in our analysis,  we propose adaptations of well-known estimators, such as kernel smoothing \cite{wand1994kernel} or machine learning approaches such as kernel ridge regression \cite{vovk2013kernel}, that accommodate the survey design structure as well as complex objects like the proposed physical activity distribution representation of accelerometer data. Including these nonparametric regression models can be a valuable option for practitioners to utilize their data more effectively by considering the study design's specific nature. Although there does not exist a vast amount of literature on estimation of nonparametric regression using survey data \cite{harms2010kernel, lumley2017fitting}, such techniques have the potential to contribute to obtain new clinical findings by modeling complex data relations that are common in biology and related fields. Since survey data can lead to more reliable conclusions than observational data \cite{lumley2011complex,ackerman2021generalizing}, the development of these tools for more complex data objects such as physical activity distributions has a high potential impact. As it it is increasingly frequent in standard clinical practice to use medical devices that monitor patient conditions with high temporal resolution, the techniques proposed in this paper can potentially be used to handle the resulting complex statistical objects quite broadly.

\subsection{Contributions}

We summarize the main contributions of the paper below.

\begin{enumerate}
	\item A new representation of accelerometer data is proposed that extends compositional data metrics in this domain to a functional context. The continuous gait component is modeled through a density function, while inactivity time is modeled as a proportion.  Importantly, this representation automatically captures compositional metrics and other summary measures that are widely used, such as total activity or proportion of inactive time, without the need to introduce any prior expert knowledge to subjectively define summary metrics.
	
	\item As the above representation is a constrained datum due to its distributional nature, the Wasserstein geometry is utlized to compare probability distributions via distances between quantile functions, similar to the glucodensity approach of \cite{matabuena2020glucodensities}.

	\item To obtain reliable conclusions from the NHANES database, the specific survey design is used to inform model fitting. To date, few if any techniques have been proposed to incorporate complex data objects such as probability distributions using survey methods. This paper expands general kernel smoothing and kernel ridge regression to estimate nonparametric regression models involving physical activity distributions with survey data.
	
	\item In the public accelerometer database NHANES:   
	
	\begin{itemize}
		\item The proposed representation, together with the inclusion of the new nonparametric survey regression models, are able to capture relevant biomarkers associated with health, such as age, body mass index (BMI), and cholesterol, better than standard metrics widely used in accelerometer field.

		\item The physical activity distributions are used in nonparametric models to predict five-year mortality, and this new representation is shown to provide additional insight compared to existing methods in terms of identifying risk patients and extreme physical inactivity profiles.

	\end{itemize}
	
\end{enumerate}

%

\subsection{Outline}
The structure of the paper is summarized as follows. First, we describe the data analyzed in the NHANES database. Next, we introduce the new representation along with notions about statistics in metric spaces.	Then, we present the approach for fitting nonparametric regression models based on survey data techniques in order to handle physical activity distributions computed from the NHANES database. Subsequently, we validate and illustrate our representation through its performance in different prediction tasks. Finally, we discuss the results obtained and the potential applications of the methodology proposed herein for other accelerometer and wearable device data.

\section{NHANES-database}\label{section:data}

The National Health and Nutrition Examination Survey (NHANES) is an extensive, stratified, multistage survey conducted by the Centers for Disease Control (CDC) that collects health and nutrition data on the US population.     

The NHANES data are publicly available from the CDC \url{https://www.cdc.gov/nchs/nhanes/index.htm}  and are broadly categorized into six areas: demographics, dietary, examination, laboratory, questionnaires, and limited access. The accelerometer data for a Particular NHANES cohort can be downloaded from the “Physical Activity Monitor” subcategory under the “Examination data” tab. 

As noted in \cite{leroux2019organizing}, manipulation of the raw NHANES accelerometer data can be complex mainly because of the large size of the data,  the lack of available software, the intricate patterns of heterogeneity of the missing data, and the need to adjust sampling weights within a specific statistical analysis.

Motivated by this problem, \cite{leroux2019organizing} introduced the R package $rnhanesdata$ in which i) the accelerometer data and other essential variables of the patients, such as age, sex, comorbidities, life-styles, and time of mortality, are organized;  and ii) the function $reweightaccel$ calculates the sample survey weights. Additionally, the authors' codes from their statistical analyses used to predict five-year mortality from the physical activity patterns and other variables discussed above have been made available on GitHub \url{https://andrew-leroux.github.io/rnhanesdata/articles/vignette_prediction.html}.

In this work, a subset of patients that are between $68$ and $85$ years old will be used. This subset is different than that employed by \cite{leroux2019organizing} that involve patients within a wider age range ($50$-$85$ years old). The decision to restrict to a narrower age range in our applications was made for two reasons: first, although the Area Under Curve (AUC) metric was high in predicting five-year mortality, the predictive models fitted in \cite{leroux2019organizing} does not classify any individuals as dead, partially motivated due to the large imbalance between classes in this data set. As  a consequence,  the classical sensitivity vs. specifity analysis that is captured by the ROC curve is not sensible, and AUC may not be the best metric to assess the usefulness and predictive capacity of a clinical diagnostic model in this type of supervised modelling. Second, we think it is more clear to constrain the analysis to a more specific target population that can show more realistically the impact of physical inactivity than a more general and heterogeneous sample that involves lower-risk patients. At the same time, one must interpret the impact of physical activity on mortality with caution as opposed to the blind use of standard model performance metrics. For example, in this domain, if a model incorrectly predicts that a patient will die, any negative impact may be very minor, with potential positive impacts such as identifying a high-risk individual who may be able to transform their lifestyle in five years, reversing their medical condition. While we cannot hope to predict mortality using only physical activity levels, these tools can serve as instruments to identify highly inactive patients phenotypes with a more urgent need for physical activity programs defined according to their specific characteristics. In general, models used to predict mortality in five years have limited predictive capacity (see for example \cite{Griffith1724, ganna20155}), as it is likely necessary to use longitudinal models that realistically capture dynamic health evolution in this type of predictive task \cite{tsiatis2019dynamic}.

\begin{figure}[ht!]
	\centering
	\includegraphics[width=0.7\linewidth]{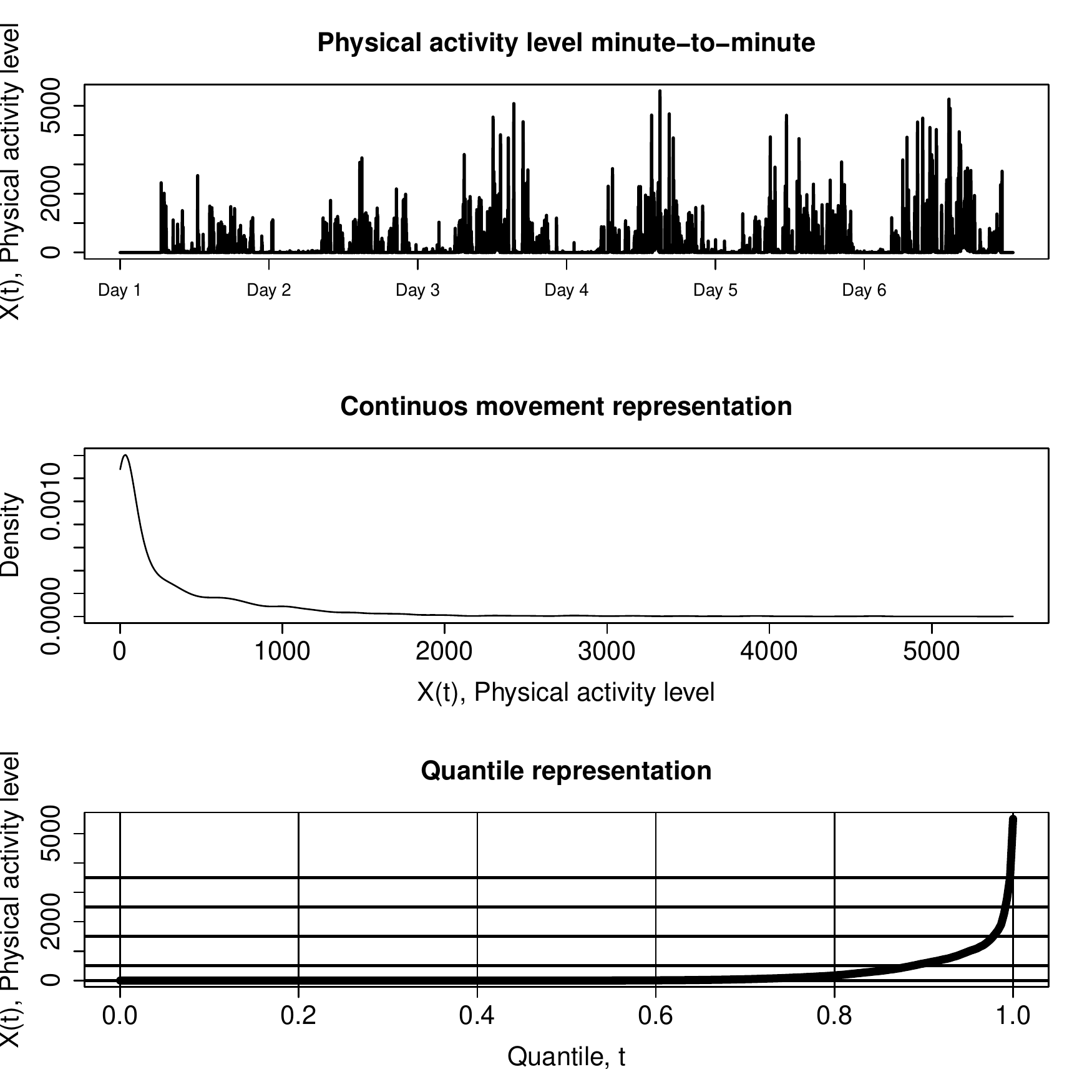}
	\caption{Example of transforming the raw accelerometer signal distributional profile for a randomly selected individual: (top) Physical activity recordings in real time; (middle) density function for active movement; and (bottom) quantile representation.}
	\label{fig:inicialpaper}
\end{figure}

\begin{table}[ht!]
	\centering
	\scalebox{0.85}{\begin{tabular}{lll}
			\hline
			Variable &  Survivors & Decedents  \\ 
			\hline
			TAC  & $193735.4$ $(115239.8)$  & $129456.5$ $(76978.72)$   \\ 
			Age  &  $72.3$ $(4.5)$  & $75.5$ $(5.5)$   \\ 
			MVPA  &  $12.3$ $(4.5)$  & $4.5$ $(9.6)$  \\ 
			ASTP  &  $0.3$ $(0.1)$  & $0.37$ $(0.12)$   \\ 
			Sedentary time  & $1126.7$ $(112.1)$  & $1180$ $(110.5)$  \\ 
			TLAC  &  $2651.7$ $(764.3)$  & $2328.7$ $(766.7)$  \\ 
			Mobility problem  & $367$ $(37\%)$  & $129$ $(60.0\%)$  \\ 
			SATP  &  $0.078$ $(0.02)$  & $0.075$ $(0.02)$  \\ 
			\textbf{Education}  &    &  \\ 
			Less than high school & $393$ $(39 \%)$   & $81$ $(38\%)$  \\ 
			High school  &  $262$ $(26\%)$  & $64$ $(30\%)$   \\ 
			More than high school & $349$   $(35\%)$  & $71$ $(33\%)$   \\ 
			\textbf{Drinking Status}  &  & \\ 
			Moderate Drinker  & $442$ $(44\%)$   & $75$ $(35\%)$   \\ 
			Non-Drinker   & $496$ $(49\%)$   &  $118$  $(55)$  \\ 
			Heavy Drinker  & $40$ $(3\%)$   & $15$ $(7\%)$   \\ 
			Missing alcohol  &  $26$ $(2\%)$ & $8$ $(4\%)$   \\ 
			\textbf{Smoking Status}  &  &  \\ 
			Never  & $460$ $(46\%)$  & $68$ $(13\%)$   \\ 
			Former  & $466$ $(46\%)$  & $110$ $(51\%)$    \\ 
			Current  & $78$ $(8\%)$   &  $38$ $(18\%)$ \\ 
			CHF  & $115$ $(11\%)$ & $37$ $(17\%)$ \\ 
			\textbf{Gender}  &  &  \\ 
			Male & $519$ $(52\%)$   & $145$ $(67\%)$   \\ 
			Female & $485$ $(48\%)$ &  $71$ $34\%$ \\ 
			Diabetes & $163$ $(18\%)$    & $48$ $(22\%)$  \\ 
			Cancer  & $231$ $23\%$   & $57$ $26\%$ \\ 
			\textbf{BMI}  &  &  \\ 
			Normal  & $283$ $28\%$  & $75$ $(35\%)$  \\ 
			Underweight  & $7$ $1\%$  & $7$ $(3\%)$   \\ 
			Overweight  & $414$ $41\%$   &  $76$ $(35\%)$   \\ 
			Obese  & $300$ $(29\%)$   &  $58$ $(27\%)$    \\
			CHD  & $115$ $(11\%)$  & $37$ $(17\%)$  \\ 
			Stroke  & $71$ $(7\%)$  & $31$ $(14\%)$\\ 
			\textbf{Race}  &  &  \\ 
			White  &  $648$ $(65 \%)$ &  $161$ $(74 \%)$ \\ 
			Mexican American  & $172$ $(17\%)$  & $22$ $(18.8\%)$    \\ 
			Other Hispanic  & $17$ $(2\%)$  & $0$ $(0\%)$    \\ 
			Black  & $144$ $(14\%)$  & $28$ $(10\%)$    \\ 
			Other  & $23$ $(2\%)$  & $5$ $(2\%)$    \\ 
			Wear time  & $878.4$ $(161.0)$  & $892.2$ $(164.98)$  \\
	\end{tabular}}
	\caption{Patients characteristic of sample considered in this work beetween live and deceased patients. The Table extracted show  means (standard deviation) or $n$ $(\%)$. In bold, multivariate categorical variables.\\ 
		TAC total activity count (TAC); total log-transformed  activity count (TLAC);   total minutes of moderate/vigorous physical activity (MVPA);  active to sedentary/sleep/non-wear transition probability (ASTP);  sedentary/sleep/non-wear to active transition probability (SATP); Coronary heart disease CHD (CHD); Congestive heart failure (CHF)}.
	\label{table:tabla1}
\end{table}

In order to explore the data, we estimate some basic patient characteristics of the patients examined using the survey weights for this target population computed using the R package $rnhanesdata$  \cite{leroux2019organizing}. Table \ref{table:tabla1} contains population characteristics of individuals according to their mortality status after five years.  Furthermore, in Figure \ref{fig:inicialpaper}, we show the raw data during several days of one participant selected randomly from the database. For each individual, while the device is worn, accelerometer devices collect an estimate of minute-by-minute energy expenditure. However, given that the device is not worn all day, the recorded signal cannot be continuous, and there are intricate missing data patterns. Precisely with this same database, several works have proposed methods of missing data to address this problem \cite{ae2018missing}. In our particular setting, we rely on the pre-processing done in \cite{leroux2019organizing}.

\section{Functional representation of accelerometer data and regression models}

\subsection{Definition of functional representation}\label{section:representacion}

First, we introduce the formal definition of the new representation. For a patient $i$, let $n_i$ be the number of observations recorded in form of pairs $(t_{ij},X_{ij})$, $j=1,\cdots,n_i$, where the $t_{ij}$ are a sequence of time points in the interval $[0,T_i]$ in which the accelerometer records activity information, and $X_{ij}$ is the measurement of the accelerometer at time $t_{ij}$. 	Unlike continuous glucose monitoring data, accelerometer readings of exactly zero are quite frequent, representing physical inactivity. Thus, in our distributional representation, we will assign positive probability mass at zero equal to the fraction of total time that the individual is physically inactive. In addition, the range of values measured by the accelerometer varies widely between individuals and groups, which can present difficulties when trying to apply common distribution data analysis methods, for example functional transformations \cite{van2014bayes, petersen2016functional, hron2016simplicial} that can be an alternative strategy to handle the representation that we specify below. 

In order to handle accelerometer data gathered over different monitoring periods in free-living environments, we propose to utilize a cumulative distribution function $F_i(x)$ for each individual.  Formally, consider a latent process $Y_i(t)$ such that the accelerometer measures $X_{ij} = Y_i(t_{ij})$ $(j=1,\ldots,n_i)$, and define $F_i$ as     
\begin{equation}\label{rep1}
F_i(x) = \frac{1}{T_i} \int_0^{T_i} \mathbf{1}(Y_i(t) \leq x)dt, \quad \text{for } x \geq 0.
\end{equation}
This definition corresponds to using $x=0$ as a cutoff for inactivity; in the NHANES data set, it always holds that $F_i(0)>0$. Thus, if $U_i$ is a random variable uniformly distributed on $[0,T_i]$ that is independent of $Y_i,$ $F_i$ is the distribution function of $Y_i(U_i).$  In practice, one could use another reasonable cutoff for inactivity.  For example, other studies have used accelerometer measures between $0-100$ to quantify the inactive range. In this case, one would define $F_i$ as the distribution of the censored random variable which takes the value $100$ whenever $Y_i(U_i) \leq 100$ and $Y_i(U_i)$ otherwise.

Analogously, we may be interested in truncating the latent process from above, for example to combine measurements representing high-intensity exercise, e.g., device observations greater than or equal to $3500$. For instance, this idea can be exploited to establish high-intensity exercise benefits in the prediction of mortality or another relevant outcome. Practically speaking, an upper threshold of this type would lead to a simpler model that could be beneficial in the predictive task. Then $F_i$ would be the distribution of the censored random variable taking values $Y_i(U_i)$ whenever this is at most $3500$, and $3500$ otherwise.  Combination of lower and upper cutoffs would be treated in a similarly straightforward manner.

In the remainder of the paper,  we define $F_i$ as in \eqref{rep1}, and denote $\mathbb{P}^{i}_{inactive}=  F_i(0)$, $F^i_{active}(x) = F_i(x) - F_i(0)$ for $x > 0,$ and $f^{i}_{active}(x)= [F^i_{active}]'(x)$. Hence, $F_i(x)=\mathbb{P}^{i}_{inactive}+\int_{0}^{x}f^{i}_{active}(s) \mathrm{d}s$, which more clearly demonstrates the mixed nature of the distribution.  In real world settings, $\mathbb{P}^{i}_{inactive}$ and $f^{i}_{active}(\cdot)$ are not observed, but must estimated from the observed sample following, which we carry out using the following two-step strategy. First, we estimate the proportion of inactivity-time, that is 
$$\hat{\mathbb{P}}^{i}_{inactive}= \frac{1}{n_i} \sum_{j=1 }^{n_i}\mathbf{1}_{\{X_{ij}=0\}}.$$
Second, we estimate the continuous physical activity profile as conditional smooth density-function.  Letting $\mathbb{K}$ denote a univariate probability density function and $h^i > 0$ the bandwidth parameter, define

\begin{equation*}
\hat{f}^{i}_{active}(x)=    (1-\hat{\mathbb{P}}^{i}_{inactive})   \frac{1}{n_i^{active}h^{i}} \sum_{j: X_{ij} > 0} \mathbb{K}\left(\frac{X_{ij}-x}{h^{i}}\right), 
\end{equation*}

where $n_i^{active}= \sum_{i=1}^{n_1}\mathbf{1}_{\{X_{ij}>0\}}$. In our experiments, the Gaussian kernel was used for $\mathbb{K}$ and the smoothing parameter was selected through Silverman's ``rule of thumb." \cite{silverman1986density}. More discussion about density estimation procedure and kernel bandwidth selection with biosensor data can be found in the glucodensity paper \cite{matabuena2020glucodensities}.

\subsection{Statistical Framework for the Distributional Representation}

While the representation of physical activity levels via the inactivity probability $\hat{\mathbb{P}}^i_{active}$ and activity density $f^i_{active}$ provides a rich and fairly comprehensive representation of the accelerometer data, the mathematical constraints of these objects makes statistical analysis challenging.  In particular, naive application of functional data analysis for the $f^i_{active}$ is known to yield results that are often difficult to interpret, as these methods do not respect the inherent constraints.  Thus, we will work under the same framework outlined in \cite{matabuena2020glucodensities} based on the Wasserstein geometry of optimal transport \cite{villani2008optimal, ambrosio2013user}. This metric has theoretical appeal, has given intuitive results in a variety of applications, and possesses many computational advantages \cite{peyre2019computational} due to its connection to quantile functions, as will be seen below.  In particular, while it is still helpful to compute the density estimates as in outlined previously in order to perform exploratory analysis and visualize results, these are not strictly necessary for the model fitting described below, where it suffices to obtain estimates of quantile functions.  Moreover, due to the mixed nature of the physical activity level distributions, the Wasserstein geometry is even more attractive as it accommodates such distributions without any special adaptation. 

Next, we define the space of the physical activity distributional representations. Let $A:= \{f:(0,\infty) \to \mathbb{R}^+: \int_{0}^\infty f(x)\mathrm{d}x<1 \text{ and } \int_{0}^{\infty}  x^2f(x)\mathrm{d}x < \infty \}$. Then the activity distributions constitute the set $\mathcal{D} \subset [0,1] \times A,$ where $(c_f, f) \in \mathcal{D}$ if $f \in \mathcal{A}$ and $c_f = 1 - \int_0^\infty f(x)\mathrm{d}x.$  Given two arbitrary inactive-active representations $\mathfrak{f}=(c_f, f)$ and $\mathfrak{g}=(c_g, g) \in \mathcal{D}$, the 2-Wasserstein (or simply Wasserstein) distance between them is
\begin{equation}
\label{eq: Wdis}
d_{W^{2}}(\mathfrak{f},\mathfrak{g})=  \sqrt{\int_{0}^{1} (F^{-1}(t)-G^{-1}(t))^{2}}\mathrm{d}t,
\end{equation}
where $F^{-1}$ and $G^{-1}$ are the quantile (inverse cumulative distribution) functions corresponding to the distributions represented by $\mathfrak{f}$ and $\mathfrak{g},$ respectively.

Given a metric or distance $d$ on $\mathcal{D},$ of which $d_{W^2}$ is one example, and a random variable $\mathfrak{f}$ defined on $\mathcal{D}$, the \emph{Fr\'echet mean} of $f$ \cite{AIHP_1948__10_4_215_0} is
\begin{equation*}
\mu_{\mathfrak{f}}= \arg \min_{g\in \mathcal{D}} E(d^{2}(\mathfrak{f},g)).
\end{equation*}
The corresponding \emph{Fr\'echet variance} of $\mathfrak{f}$ is then
\begin{equation*}
\sigma_\mathfrak{f}^{2}=  E(d^{2}(\mathfrak{f},\mu_{\mathfrak{f}})).
\end{equation*}
With the particular choice $d = d_{W^2},$ we have
\begin{equation*}
\mu_{\mathfrak{f}}= \arg \min_{\mathfrak{g}\in \mathcal{D}}  E\left[\int_{0}^{1} (F^{-1}(t)-G^{-1}(t))^{2} \mathrm{d}t\right],
\end{equation*}
and, with $Q_{\mathfrak{f}}$ denoting the quantile function corresponding to $\mu_{\mathfrak{f}},$
\begin{equation*}
\sigma_{\mathfrak{f}}^{2}= E\left[\int_{0}^{1} (F^{-1}(t)-Q_{\mathfrak{f}}(t))^{2} \mathrm{d} t\right].  
\end{equation*}

\begin{figure}[ht!]
	\centering
	\includegraphics[width=0.7\linewidth]{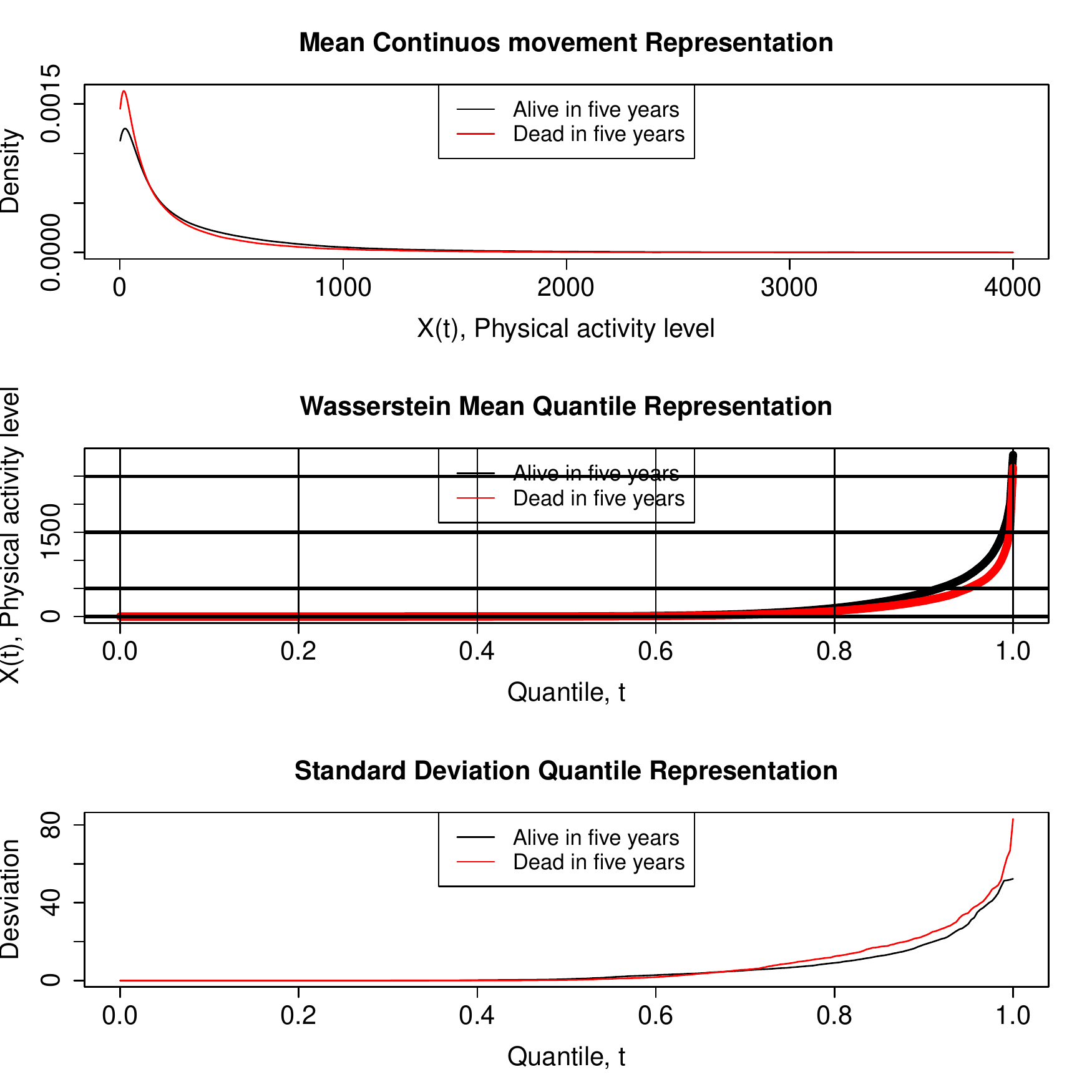}
	\caption{Summary curves of physical activity distributions in Quantile space (mean and standard deviation) between life and deceased patients groups after five years together with the mean of continuos movement representation}
	\label{fig:curvasmortalidad}
\end{figure}

Given $n$ samples of accelerometer measures belong $n$ individals $\{X_{i,j}\}_{j=1}^{n_i}$, $i=1,\dots,n$, we can form empirical quantile functions $\hat{Q}_i = \hat{F}_i^{-1}$.  Then, due to the Euclidean nature of \eqref{eq: Wdis}, the empirical Fr\'echet mean and variance, written in terms of quantile functions, take the form of
\begin{equation*}
\overline{Q}(t)= \hat{Q}_{\mathfrak{f}}(t) = \frac{1}{n} \sum_{i=1}^{n}  \hat{Q}_i(t), \hspace{0.2cm} t\in [0,1], \, \text{and}
\end{equation*}
\begin{equation*}
\hat{\sigma}_{\mathfrak{f}}^{2}= \frac{1}{n-1}\sum_{i=1}^{n}\int_{0}^{t}  (\hat{Q}_i(t)-\overline{Q}(t))^{2} \mathrm{d}t.
\end{equation*}


In this case, also we construct the estimated pointwise quantile variance curve $\tilde{\sigma}^{2}_{f}(t)$, representing the sample variance of the values $\hat{Q}_1(t),\ldots,\hat{Q}_n(t),$ for each $t\in [0,1]$.	Figure \ref{fig:inicialpaper} illustrates the process of transforming raw data into our representation. In addition, Figure \ref{fig:curvasmortalidad} shows the mean and variance curve of our representation, where patients are grouped by mortality status after five years.


\subsection{Survey regression models}
\label{survReg}

The individuals that we analyze from the NHANES database do not represent a random sample of a population. Instead, they are the result of a structured sample of a complex survey design from a finite population of $N$ individuals. In order to perform inference correctly and obtain reliable results, one must account for the effects of the specific sample design when building a predictive model. Note that the information provided by this type of survey data is typically richer than those used in most medical studies that are of an observational nature \cite{lumley2011complex, ackerman2021generalizing}. In the latter case, the research does not explicitly control the sampling mechanism, so that obtaining a representative population results can be challenging and would often demand colossal data volumes.

Suppose that observations $\{(Y_i, X_i ); i \in S\}$ are available, where $Y_i$ is a scalar response variable, and $X_i$ a collection of covariates taking values in a metric space.  The index set $S$ represents a sample of $n$ units from a finite population. To account for this sampling, each individual $i \in S$ will be associated with a positive weight $w_i$. In our analyses, these weights were taken to be the the inverse of the probability $\pi_i > 0$ of being selected into the sample \cite{kish1965survey}, i.e. $w_i=1/\pi_i,$ although more general situations are available \cite{lumley2017fitting}.  When performing estimation with survey data, a common approach is to use the $w_i$ to define weighted versions of usual estimates designed for random samples.  For example, the normalization Horvitz-Thompson estimator \cite{horvitz1952generalization, https://doi.org/10.1111/j.1467-985X.2006.00426.x} for the population average of the $Y_i$ is the weighted sample average
\begin{equation}
\label{wSampMean}
\overline{Y}_{w}= \sum_{i\in S} \frac{\frac{1}{\pi_i}Y_i }{\sum_{i\in S} \frac{1}{\pi_i}}= \sum_{i\in S} \frac{w_i Y_i}{\sum_{i\in S} w_i}.
\end{equation}

Although simplistic in appearance, the researcher often controls the study and sampling design for survey data so that weights can be calculated easily without the need to resort to any data-driven approach. In particular, the sampling weights are calculated according to some demographic variables that are essential in describing population composition. These variables are used in the process of selecting a sample of elements from a population. In the NHANES cohort, we can download for each study participant their representative weight in the United States population. However, we must re-weight according to variables (not all NHANES participants have accelerometer data) and specific patients that we introduce in the analysis (for example, subset of patients with more than 68 years).

In this paper, we propose to use a general kernel smoother \cite{wand1994kernel, chacon2018multivariate} for survey data with weights that are composed of both the sampling weights $w_i$ as well as the usual local weights that appear in such kernel methods. One main advantage of this estimator is its flexibility, as it is valid for either regression or classification problems. In addition, we also extend kernel ridge-regression \cite{vovk2013kernel}, but this method is only appropriate for a continuous response variable. Theoretical properties such as asymptotic results and convergence rates, although interesting, will not be investigated here as the main goal is to solve and model a real problem.

\subsubsection{Kernel smoother for survey data.}

Suppose the mean regression model
\begin{equation}\label{reg1}
Y= m(X)+\epsilon
\end{equation}
holds, where $\epsilon$ is a random error term satisfying $E(\epsilon|X)=0$.  Hence, the value $m(X)$ represents the conditional mean of $Y$ given $X,$ where $m$ is assumed to be a smooth function.  Given a sample $\{(Y_i, X_i, w_i); i \in S\}$ of size $n$ from the finite population as described above, an estimate of $m(x)$ for a generic input $x$ may be obtained using the standard kernel estimator \cite{wand1994kernel}
\begin{equation}\label{smooth1}
\hat{m}(x)= \sum_{i\in S}^{} s(X_i,x) Y_i, 
\end{equation}
where $s(X_i,x)$ is an appropriate weight function that provides more weight for predictors $X_i$ with smaller distance to $x$.  Furthermore, the constraint $\sum_{i\in s}^{} s(X_i,x)=1$ must be satisfied for all $x$ to obtain a coherent estimator.  Typical choices for $s$ include Nadayara-Watson weights $s(X_i,x)= \frac{\mathbb{K}(  h^{-1}d(X_i,x))}{\sum_{i\in s}^{} \mathbb{K}(h^{-1}d(X_i,x))}$, where $d$ is a metric on the set of predictors, for example the Wasserstein distance defined in \eqref{eq: Wdis} if the covariate $X$ represents a physical activity level distribution, and $h>0$ is the smoothing-parameter.  The generalization of the standard Nadaraya-Watson estimator, which was originally proposed for scalar or vector predictors, to more abstract data types has been used to handle functional predictors \cite{ferraty2006nonparametric} as well as predictors and responses in more general spaces that possess a metric \cite{steinke2010nonparametric}.  Here, we will utilize the quantile functional representation of the accelerometer data along with the Wasserstein metric to incorporate these complex objects as predictors of relevant outcomes.

Due to the survey design, we make the necessary adjustment to the usual Nadaraya-Watson weights by scaling them according to the survey weights $w_i.$  Specifically, we set 
$s(X_i,x)= \frac{\mathbb{K}(h^{-1}d(X_i,x)) w_i}{\sum_{i\in s}^{} \mathbb{K}(h^{-1}d(X_i,x)) w_i} $.  This definition reflects that an observation should be given higher weight when the probability of selection is lower (large values of $w_i$), consistent with the principles outlined in \cite{horvitz1952generalization}, and when the observed input $X_i$ is closer to the input $x$ at which one desires an estimate of the conditional mean.  In general, the kernel smoother in \eqref{smooth1} corresponds to a convex combination of the observed responses, a property that is not shared by similar smoothers, for example local linear regression estimators. 

An important consequence of this convexity property is in the case of a binary response variable $Y\in \{0,1\}.$ In this case, $m(x) \in [0,1]$ represents a probability, so that \eqref{smooth1} yields an estimate $\hat{m}(x) \in [0,1]$ that can be used in classification tasks, for example, without any post hoc modification.  
When $Y$ represents a categorical variable that can assume more than two values, a simple modification of \eqref{smooth1} can still be used to produce valid estimates of the various probabilities.

\subsubsection{Kernel ridge regression for survey data.}\label{section:ridge} 


Let $X \in \mathcal{D}$ be an object of complex type as our probability distribution of physical activity levels. The Reproducing Kernel Hilbert Space (RKHS) learning paradigm \cite{rakotomamonjy2005frames} provides a unique and rich framework to create new and more flexible predictive models that can handle abstract variables $X$ as predictors by assuming that the regression function $m$ in \eqref{reg1} is an element of a space of functions $V$ on $\mathcal{D}$ that is an RKHS. This section focuses attention on a method known as kernel ridge regression that leverages the properties of an RKHS to produce estimates that can be viewed as generalizations of the usual ridge regression estimator for linear models. In the following, we summarize the necessary components of the RKHS-based model and its estimator, and then adapt the estimator to the case of survey data.

For each input value $x \in \mathcal{D}$, 
one way of defining an RKHS is to begin with a kernel $K: \mathcal{D} \times \mathcal{D} \rightarrow \mathbb{R}$ that is symmetric and positive definite.  Beginning with functions of the type $\phi_x(\cdot) = K(x,\cdot)$ as basic elements, one can construct a Hilbert space of functions by taking linear combinations and, finally, by taking the usual metric completion.  The constructed Hilbert space $V$ can be shown to have the inner product $\langle \cdot, \cdot\rangle_V$ with the property that $\langle \phi(x), \phi(y) \rangle_V= K(x,y)$.  Furthermore, for any $f \in V,$ one has $f(x) = \langle \phi_x, f\rangle_V = \langle K(x,\cdot), f\rangle_V,$ so that $K$ is often referred to as a reproducing kernel, or the kernel that generates $V.$   Observe that the use of the term kernel is distinct from that of the previous section. For clarity, a distinct notation has been introduced for the bivariate kernel of the current section. 

Considering the model defined Equation $\ref{reg1}$, an alternative to the smoothing method of the previous section is to assume that the regression function $m(\cdot)\in V$.  Given the infinite-dimensional nature of $V$, estimation of $m(\cdot)$ through the use of least squares, i.e.
\begin{equation}\label{eqn:optimizationprimero}
\hat{m}= \arg \min_{m\in V} \sum_{i\in S}^{} (Y_i- m(X_i))^{2}+\lambda  \norm{m}^{2}
\end{equation} 
is ill-defined, where $\norm{\cdot}$ is the Hilbertian norm on $V.$  Specifically, there are many different solutions to \eqref{eqn:optimizationprimero} that attain zero empirical error.  Naturally, overfitting the model in this way results in poor predictive capacity for new observations.

In the RKHS framework, it is common to introduce a norm-based penalty on $m$ in the optimization procedure to induce regularization.  The kernel ridge regression estimator then becomes
\begin{equation}\label{eqn:optimization}
\hat{m}= \arg \min_{m\in V} \sum_{i\in S}^{} (Y_i- m(X_i))^{2}+\lambda  \norm{m}^{2},
\end{equation}
where $\lambda$ is the regularization parameter that controls the usual trade-off between bias and variance, which in turn determines the capacity of the model to generalize to new observations.

By the classical Representer Theorem \cite{scholkopf2001generalized}, the solution to \eqref{eqn:optimization} is known to take the form $\hat{m}(\cdot)= \sum_{i\in S} \alpha_i K(\cdot, X_i)$, so that the estimator is a linear combination of the kernel features $K(\cdot, X_i)$ with coefficients $\alpha_i$. Solving \eqref{eqn:optimization} under this restricted form of $m$ results in the coefficient estimates $\hat{\alpha}= (K+\lambda I)^{-1}Y$, where 

$$  K=  \begin{pmatrix} K(X_1,X_1), \dots, K(X_1,X_n) \\ K(X_2,X_1), \dots, K(X_2,X_n) \\ \cdots \\ K(X_n, X_1), \dots, K(X_n, X_n)  \\    \end{pmatrix}, \hspace{0.2cm} Y=\begin{bmatrix}
Y_{1} \\
\vdots  \\
Y_n 
\end{bmatrix},$$
and $I$ is the $n\times n$ identity  matrix. 


To introduce survey design in the optimization problem, we can use the Horvitz–Thompson version of the estimator, namely 
\begin{equation}\label{eqn:optimization2}
\hat{m}= \arg \min_{m\in V} \sum_{i\in S}^{} w_i(Y_i- m(X_i))^{2}+\lambda  \norm{f}^{2}.
\end{equation}  
As \eqref{eqn:optimization2} remains a convex objective function, the Representer Theorem holds and the solution will retain the same structure, $\hat{m}(\cdot)= \sum_{i\in S} \alpha_i K(\cdot, X_i)$. However, the coefficients take the form of regularized weighted least squares estimates $\hat{\alpha}= (WK+\lambda I)^{-1} WY$, with $W$ being a diagonal matrix with the weights $w_i$ constituting the diagonal elements.

A notable advantage of the kernel ridge regression is that it preserves some computational advantages of linear models, as the optimal solution can be calculated via weighted least squares. This fact also simplifies selection of the tuning parameter $\lambda,$ for example using Leave One Out Cross-Validation (LOOCV), given that explicit leave-one-out formulas are available for linear estimators \cite{golub1979generalized}. 

Besides the regularization parameter, a crucial choice that determines the model's empirical performance is that of the RKHS learning space $V$ or, equivalently, the kernel $K$. In our preliminary test, the best results are obtained with the Laplace kernel, so in the following, we consider only this kernel.  Let $x$ and $y$ be two points belonging to input space, which we assume has an associated norm $||\cdot ||.$  Then the Laplacian kernel $K(x,y)$ is
\begin{equation}\label{eqn:kernellaplacian}
K(x,y)= e^{-\norm{x-y}/\sigma},
\end{equation}   
where $\sigma>0$ is a scale parameter that can be chosen heuristically as \cite{garreau2017large}
\begin{equation}
\hat{\sigma}= \sqrt{ \textrm{median}\{||X_i-X_j||^{2}:1\leq i<j\leq n\}},
\end{equation}

being in this setting,  $median(\{||X_i-X_j||^{2}:1\leq i<j\leq n\})=  \inf\{x\in \mathbb{R}: \hat{F}_{n,w}(x)\geq 0.5\}$,   with $\hat{F}_{n,w}(x)=  \sum_{1\leq i<j\leq n} w_i w_j \mathbf{1}\{||X_i-X_j||^{2}\leq x\}/ \sum_{1\leq i<j\leq n}^{} w_i w_j$, that is, we estimate the median  introducing sampling mechanism through survey weights.

%
%
%
%

\section{Results}

\subsection{Outline of Analysis Performed}

To show the potential of the new representation of accelerometer data and the application of new nonparametric survey regression estimators, the analyses described below were performed with the sample described in Section \ref{section:data}.

\begin{enumerate}
	\item Four different response variables were analyzed, namely age, Body Mass Index (BMI), blood pressure, and cholesterol levels. Kernel ridge regression was used to compare our functional representation with total activity count (TAC), the most relevant and commonly used physical variable in different studies with this database \cite{leroux2019organizing, Wsmirnova2019predictive}. 
	\item The ability of our functional representation to predict five-year mortality was assessed with the Nadayara-Watson survey estimator. Furthermore, an in-depth clinical analysis of the results provided by the algorithm is given. 
	
\end{enumerate}

The classical TAC metric described in Table \ref{table:tabla1} is entirely captured by our representation 
because it can be calculated directly from the distribution of physical activity levels.  Thus, we do not fit any statistical models as the relationship is deterministic. To simplify the comparison with standard summary metrics, we examine the gains of our representation only against the TAC metric, since this has been shown to be the most relevant physical activity variable in other studies using the NHANES database (see for example \cite{leroux2019organizing}). 

Additionally, a comparison with the usual finite-dimensional compositional metrics is not given since it requires a prior subjective definition of specific cut-offs that are usually adapted to the target population. In addition, the information is automatically registered by our representation, and preliminary experiments demonstrated that the compositional metrics were inferior to the complete distributional representation.

\subsection{Comparison of Distributional Representation with TAC}

This section shows that our representation contains more valuable information about patient health than the standard TAC metric, which has been shown to be the most relevant physical activity variable in the NHANES database.  Indeed, it can be seen as an integral part of our distributional representation; that is, this variable contains information about total energy expenditure. As the number of days that patients are monitored is different, we normalize the estimation obtained of the TAC metric by the number of days that the patient is registered.  One way to illustrate the benefits of our method is to compare the ability of each to capture essential biomarkers associated with the health and decline of physiological function. For this purpose, we select Age, Body mass index (BMI), blood pressure, and cholesterol as response variables. As a regression estimator, we select the kernel ridge-regression introduced in Section ``Kernel ridge regression for survey data''
, with the Laplacian Kernel in \eqref{eqn:kernellaplacian}. In constructing this kernel, we employ the Wasserstein distance when the predictors are distributions (which can be expressed as a norm via the quantile representation), and the Manhattan or $\ell_1$ distance in the case of scalar TAC predictors. To make a conservative assessment of physical activity levels, we calculate a survey-weighted leave-one-out version of $R$-square, defined as

\begin{equation}
\label{RsqSurvey}
R^{2} = 1- \frac{\sum_{i\in S}^{} w_i(Y_i-\hat{f}^{-i}(X_i))^{2}}{\sum_{i\in S}^{} w_i(Y_i-\overline{Y}_w)^{2}}, 
\end{equation}

where $w_i$'s are the survey weights, $\overline{Y}_w$ is defined in \eqref{wSampMean}, and $\hat{f}^{-i}(\cdot)$ is a generic regression estimate obtained after deleting the $i$-th observation. As the models are nonlinear and leave-one-out estimators are used, $R^2$ as defined in \eqref{RsqSurvey} can be negative, as seen for blood pressure as response with TAC as predictors in Table \ref{tabla:reg}, where all results are compiled.

These results demonstrate that the predictive capacity is low to moderate for all models, with age and BMI being the most predictable variables. In all cases, it is clear that our repesentation outperforms the summary metric TAC.  As the new representation retains more information than summary metrics, we hypothesize that the advantages of the former may be even more significant in larger databases. 

\begin{table}[ht!]
	\label{table:res1}
	\begin{center}
		
		\begin{tabular}{|c|c|c|}
			\hline
			& New representation & TAC \\	
			\hline
			Age	& $0.15$ & $0.07$  \\
			\hline
			BMI	& $0.05$ &  $0.01$ \\
			\hline
			Blood pressure 	& $0.02$  &  $-0.01$ \\
			\hline
			Cholesterol total & $0.034$	& $0.016$    \\
			\hline
		\end{tabular}
	\end{center}
	\caption{R-square obtained with each representation used in Kernel Ridge-Regression models with  continuous variables examined}
	\label{tabla:reg}
\end{table}

\subsection{Prediction of five-year mortality}
\label{ss:mort}

A Nadayara-Watson estimator was used to predict five-year mortality with both our representation and the TAC metric. Once again, leave-one-out prediction error was used to quantify prediction accuracy to avoid over-fitting. According to the solution selected,  $23$ patients were classified correctly with our representation, while $36$ patients who were predicted to die within five years survived. Using the TAC metric, the model predicted that all patients would die. To appreciate the practical clinical interpretation of this analysis, consider the following two patient groups:
\begin{itemize}
	\item A (risk-group): Patients that were predicted to die but survived.
	\item B (non-risk group): Patients that survived and are corrected classified by the model.
\end{itemize}

\begin{figure}[ht!]
	\centering
	\includegraphics[width=0.7\linewidth]{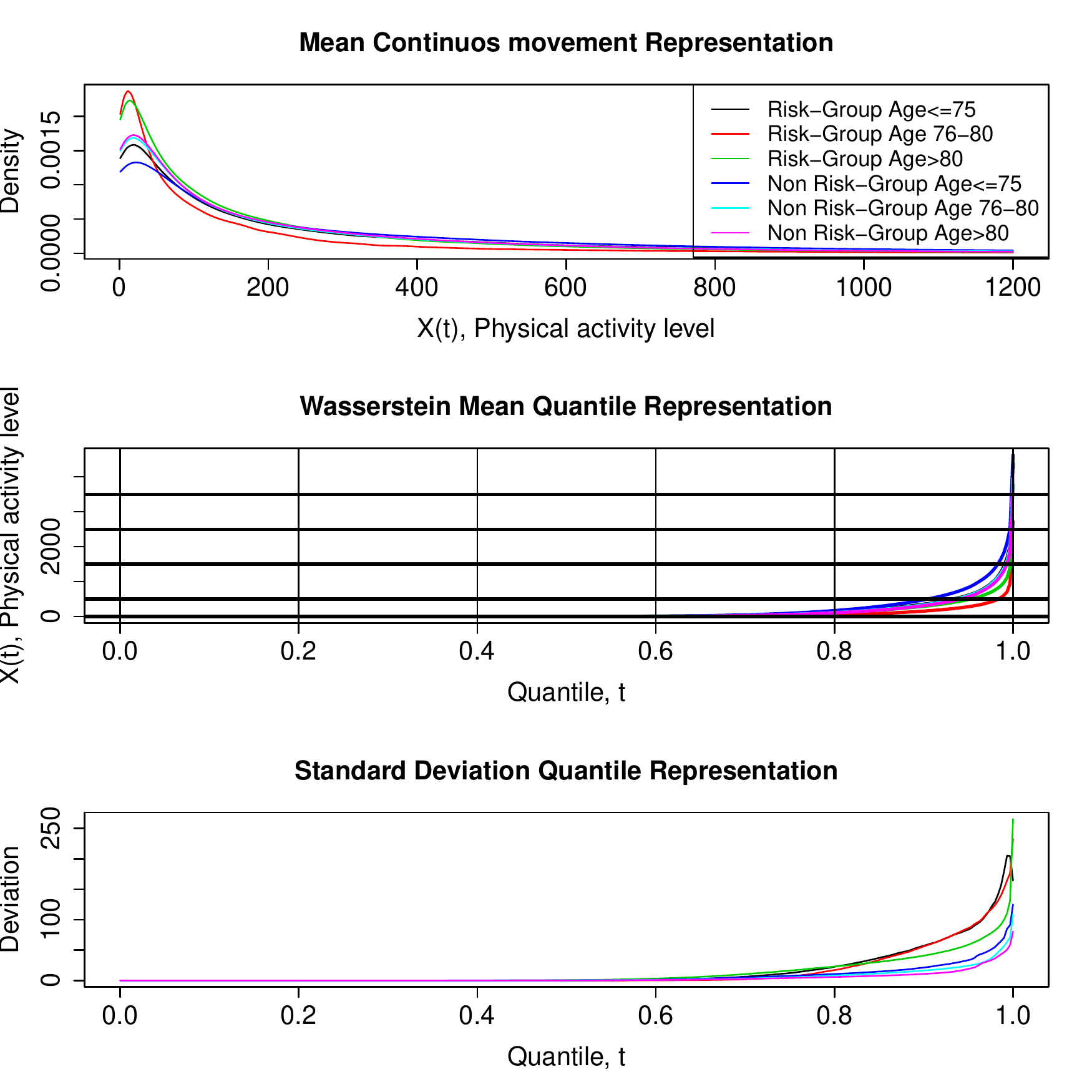}
	\caption{Wassertein Mean functional profiles (top, density of active movement; middle, quantile represtionation) and pointwise quantile standard deviation curves (bottom) of different risk and age strata.}
	\label{fig:gruposedad}
\end{figure}
\begin{figure}[ht!]
	\centering
	\includegraphics[width=0.7\linewidth]{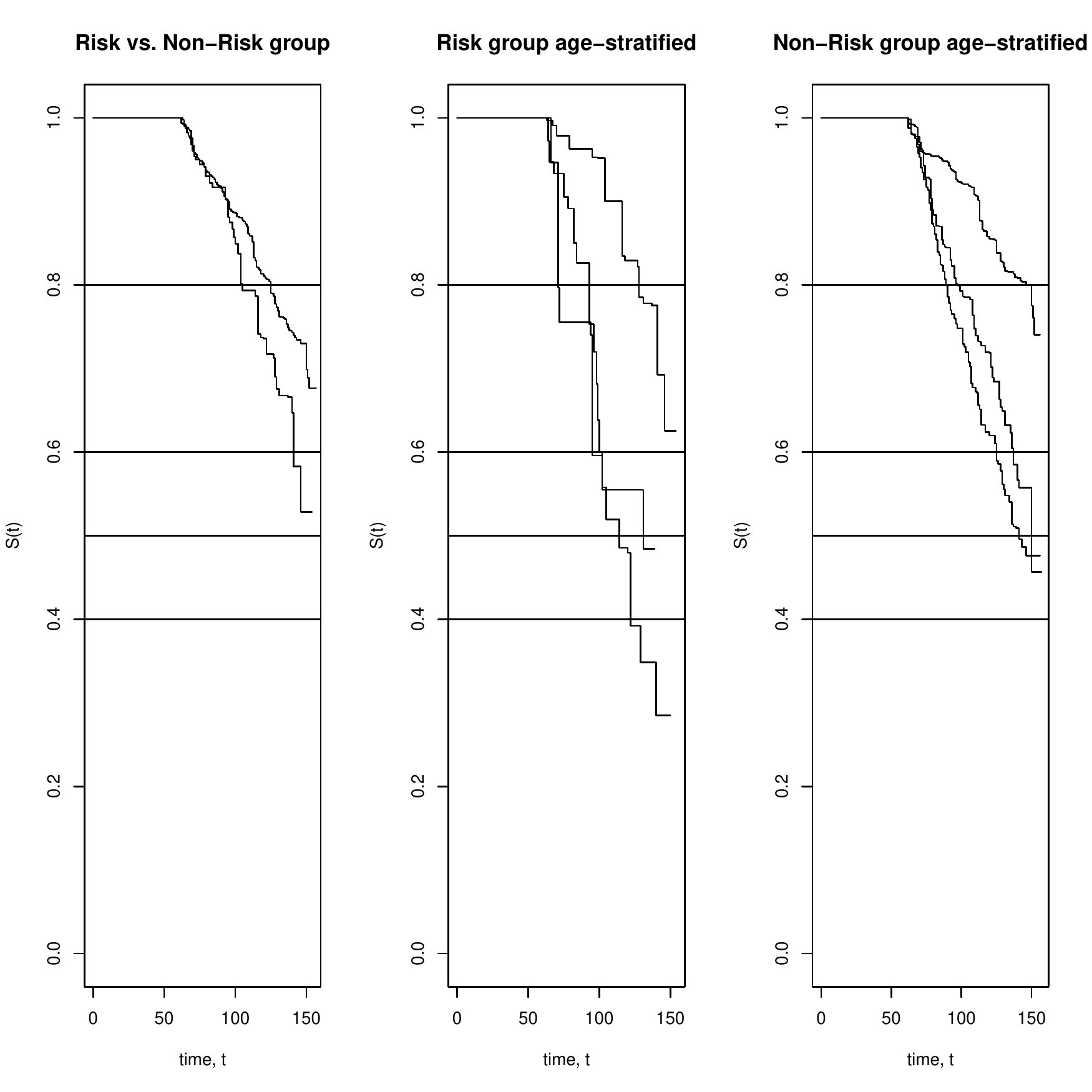}
	\caption{Survival curves of risk and non-risk group, in the entire cohort (left) and by age stratification (middle and right).}
	\label{fig:surv2}
\end{figure}

Additionally, consider the following age stratification: $68-75$, $76-80$, and $81-85$ years old.  Figure \ref{fig:gruposedad} show apparent differences in physical activity distribution between subjects in different risk groups and distinct age strata. In addition, the survival curves in Figure~\ref{fig:surv2} are lower for the risk group compared to non-risk patients, and also for more elderly patients within each of the risk and non-risk groups.  

\begin{figure}[ht!]
	\centering
	\includegraphics[width=0.7\linewidth]{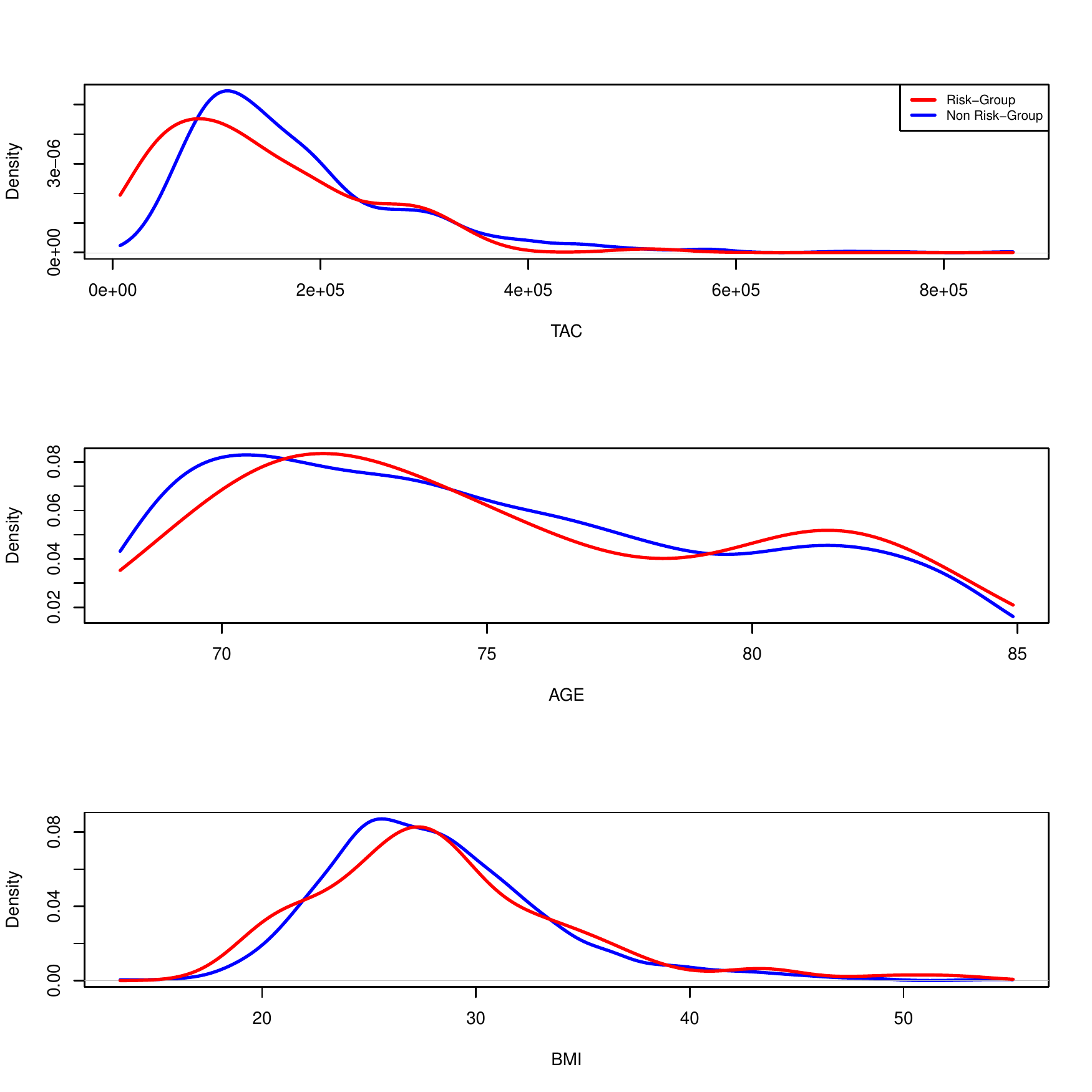}
	\caption{Density functions estimation of TAC, Age, and BMI, all important health indicators, within risk and non-risk groups.}.
	\label{fig:superhistogramas}
\end{figure}

\begin{figure}[ht!]
	\centering
	\includegraphics[width=0.7\linewidth]{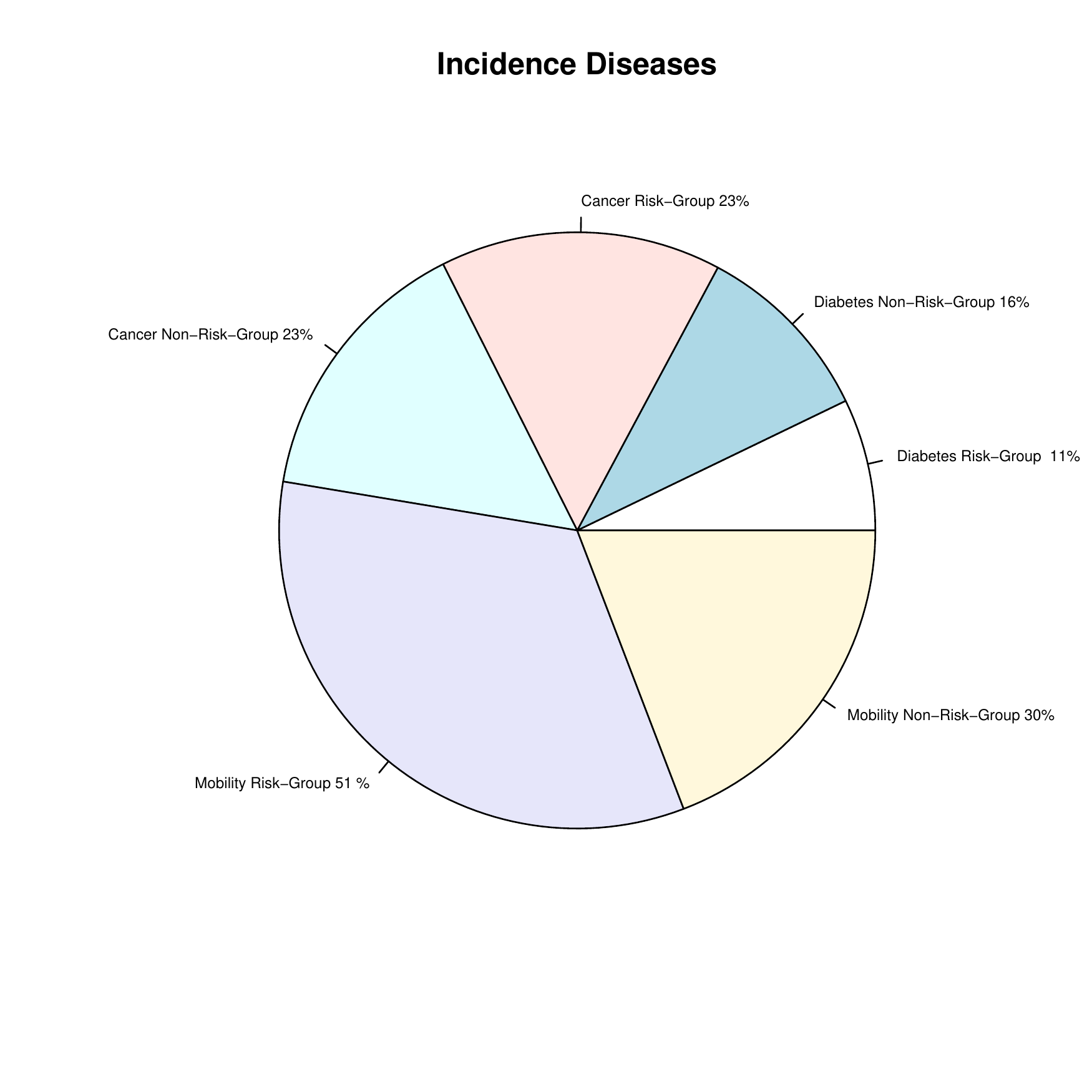}
	\caption{Distributional composition of some diseases related with mortality in risk and non-risk groups}
	\label{fig:graficoanalisisresultadoscategoricos2}
\end{figure}

Finally, we performed complementary analyses of clinical groups defined in Figures \ref{fig:superhistogramas} and \ref{fig:graficoanalisisresultadoscategoricos2} with respect to other health indicators.  In these analyses, a slightly percentage of patients older than $75$ years and with significantly higher-body mass levels belong to the non-risk groups. In addition, the percentage of diabetes patients is higher in non-risk groups. These comorbidities at least partially explain some of the incorrect classifications of our model.

\newpage

\section{Discussion}

In this work, we have introduced a new functional profile of an individual's physical activity to quantify more comprehensively the 
energy expenditure over a given period among a group of individuals monitored in free-living conditions. Our procedure can be seen as a functional extension of the so-called compositional metrics that constitute the most popular approach to date in the accelerometer field. A fundamental advantage of the new method is that it automatically captures information from compositional metrics, regardless of the cutoffs used to define them. As such, one loses no information compared to these metrics, and simultaneously avoids the need to define a-priori different cut-off points \cite{biagi2019individual,dumuid2018compositional}.  Even with expert knowledge, such a selection introduces subjectivity into the analysis, with the cutoffs inevitably depending on sample characteristics or the analysis task at hand.

In the different regression tasks considered in this paper, we have seen that the new representation shows stronger predictive power than other common summary measures such as TAC, which has been shown to be the most successful variable, for example, in predicting $5$-year mortality in the NHANES database in other studies \cite{leroux2019organizing, Wsmirnova2019predictive}. Overall, the different models assessing the associations between our representation and clinically relevant covariates, as quantified by the leave-one-out $R$-square metric, is moderate (perhaps with the exception of age), indicating that there is a large amount of variability in many biomarkers associated with patient health, such as cholesterol, blood pressure, or BMI, that is not associated with how individual physical activity levels are distributed \cite{atienza2011self, luke2011assessing}. Several epidemiological studies have used multivariate regression models with these biomarkers as response variables and summary measures of physical activity as covariates.  Although some have found physical activity to be a statistically significant factor, many of these models only assess the statistical significance of the variable, rather than its practical significance or predictive capacity \cite{pepe2004limitations}.  Assessing only statistical significance fails to accurately assess the magnitude of impact physical activity has on biomarker prediction, and obscures the conclusions and reproducibility of findings reached.

The evaluation of a clinical diagnostic model that aims to predict $5$-year mortality and includes physical activity levels as a covariate is not straightforward. We cannot hope with physical activity patterns alone to predict whether a patient will die or not. However, we can identify subjects who are at risk of disease or have a higher risk of death because of their inactivity. From this point of view, the commonly used AUC metric may not be the most adequate one. For instance, there is a critical imbalance problem, and false positives do not necessarily have a negative connotation in this task.

In this predictive model, we have also performed a clinical validation based on the longitudinal evaluation of the two groups of patients whose outcome can
not match the one predicted by our model. In particular, we have considered a group of at-risk patients and a group of non-risk patients. Risk subjects are defined as those predicted by the model to die but who survive in reality. In contrast, non-risk subjects are those who survive and are correctly classified by the model. As we can see in Section ``Prediction of five-year mortality.", despite the fact that there are lower percentages of both diabetes and  elderly patients in the at-risk group, survival is lower in that group compared to non-risk patients.  Moreover, this finding persists after stratifying by age into three categories. This fact demonstrates the effectiveness and clinical sensitivity of our representation and the physical activity information provided by accelerometer. Furthermore, a model including only the TAC metric does not provide any relevant information in clinical decision support. It cannot detect any patients who die.

Despite the large number of studies that have analyzed the impact of physical activity in the NHANES cohort against different biomarkers associated with health or with mortality and survival, few studies incorporate the complex survey mechanism in the analyses, which is very crucial for obtaining reliable and reproducible results. In fact, our preliminary analyses confirmed that, without adapting the estimators according to the specific sampling design effect, the results are entirely different (the same issue are illustrated in NHANES tutorials \url{https://wwwn.cdc.gov/nchs/nhanes/tutorials/module3.aspx}).  The correct analysis of survey data can yield more robust findings than those obtained with observational data. To increase reliability of the analyses, we model relationships between the covariates and response variable nonparametrically.  To the best of our knowledge, this work is the first to combine survey methods for the estimation of nonparametric regression models involving complex predictors such as the physical activity distributions we consider.  
In particular, we have demonstrated how to implement the Nadaraya-Watson kernel smoother as well as kernel ridge regression models in this context.

We believe that the proposed model constitutes an important step forward in the use of complex objects with this type of data, which appear naturally in some of the world's most important physical activity cohorts, to obtain data representative of the physical activity patterns of a population, and not only in NHANES as a particular case. It is likely that with the technological revolution in which we live, the use of biosensors and smart-phones in surveys in digital medicine will become increasingly common to characterize the population health \cite{10.1093/jssam/smz060} and, in particular, the physical activity levels of a population.

Currently, one of the main hot topics in the field of physical activity and exercise science is to establish the impact of high-intensity exercise on patient mortality and lifespan \cite{ekelund2019dose}. Following the methodology described in Section ``Definition of functional representation" of truncate thought specific threshold Quantile physical activity representation, we have found that the high-intensity physical pattern analyzed may not result in noteworthy additional benefits to reduce mortality. Recently, a randomized clinical trial has reached similar conclusions \cite{stensvold2020effect}. However, other observational studies have established the essential benefits of this mode of exercise \cite{gill2020linking}. Therefore, we believe that the NHANES data with a complex survey design may provide a unique source of information for answering these relevant questions more reliably than other studies with less ideal design studies such as that of the UK-Biobank \cite{leroux2020quantifying, strain2020wearable}.

One of the main reasons used to justify high-intensity exercise in this context is that it increases maximal-oxygen consumption more significantly \cite{milanovic2015effectiveness} than lower intensity exercise; in more prolonged exercise therapies such as endurance training. Maximal oxygen consumption is an essential physiological variable related to the prognosis and health of patients \cite{matabuena20186, mikkelsen2020improvement,Hoppelerjeb164327}. However, it has been shown that, in diseased and elderly populations, regular exercise increases fitness but does not necessarily increase this cardiovascular maximal parameter value \cite{Hoppelerjeb164327}, as there exists massive inter-individual variability in exercise response. In this sense, more personalized training programs and refined patient evaluation are required for the therapy to be successful \cite{matabuena2019prediction, buford2013toward}.  In \cite{matabuena20186}, to provide a solution for the problem of assessing physical condition in a straightforward and accessible way, 
using complex statistical techniques based on functional data analysis, a sub-maximal test was proposed to estimate maximal oxygen consumption accurately that only requires a minimal intensity exercise test. However, its implementation with clinical populations is still an open problem, and potential solutions and protocol test adaptations for elderly and chronic patients that compose clinical target populations are discussed in \cite{matabuena2019prediction}.

We believe that to quantify the beneficiary of physical activity in healthy and specific diseases in a more accurate way; we must employ different sample extraction and statistic analysis strategies. For example,  
instead of trying to predict physical exercise association in subjects monitored in free-living conditions with some outcomes that impact health, such as BMI. Maybe it is better to carry on clinical studies with a good design of experiments,  where we plan a series of specific patient interventions based on different long-term training programs, to assess in a more realistic way the effects in different groups of patients that have expressive clinical meaning. We think that how these patients' health therapies impact relevant outcomes such as weight or glucose values should be analyzed longitudinally in more controlled environments. To pursue this aim, it would be interesting to use a randomized clinical trial. However, perhaps because of cost and operational reasons, it might make more sense to use other study designs or even dynamic clinical trials following the SMART methodology or similar \cite{kosorok2015adaptive}, which can be supported by machine learning techniques for dynamic patient treatment assignment, even with observational data, to handle more extensive patient databases. In this setting, the proposed representation would continue to measure and represent variations in physical activity levels over a period given more comprehensively than existing methods. Moreover, we think our method will present more pronounced clinical sensitivity than the analysis performed here, in which patients showed very heterogeneous physical activity patterns as they were monitored in free-living conditions.

The use of digital medicine \cite{Topoleaaw7610} strategies using biosensors may lead to improved disease management, diagnosis, and prescription. For the advancement of this area, together with precision medicine, it is fundamental to develop a new statistical methodology to handle the complex data that registers healthy patients, such as the functional profiles of physical activity proposed here, and which can be used with data from other biosensors, wearables or smart-phones to better register patient health \cite{10.1371/journal.pbio.2001402}. Undoubtedly, this is a fascinating field of research, and the new health care revolution has only just begun; according to \cite{DORSEY2020859}, in coming years, telemedicine will \emph{"simply become medicine"}.

\section*{Data Availability}
The raw data used in this research is public and can be downloaded at \url{https://www.cdc.gov/nchs/nhanes/index.htm}. Data filtering are performed following the analysis \cite{leroux2019organizing}, and the scripts used for these purpose are freely avalaible at  \url{https://andrew-leroux.github.io/rnhanesdata/articles/vignette_prediction.html}. 
\section*{Competing Interests}
The authors declare no competing interests.

\section*{Acknowledgements}

Marcos Matabuena thanks Thomas Lumley their email answer about different questions for using the survey R package and other general questions about survey regression modeling. Also, the authors thank Andrew Leroux for their response about further questions of their accelerometer papers and rnhanesdata R package.

This work has received financial support from  National Science Foundation, Grant/Award Number: DMS-1811888;  the Spanish Ministry of Science, Innovation and Universities under Grant RTI2018-099646-B-I00, the Consellería de
Educación, Universidade e Formación Profesional and the European Regional Development Fund under Grant ED431G-2019/04.

\end{document}